\RequirePackage[2020-02-02]{latexrelease}
\documentclass[aps,reprint,amsmath,amssymb,pre,onecolumn,12pt]{revtex4-2}
\usepackage{graphicx}
\usepackage{dcolumn}
\usepackage{hyperref}
	\usepackage{mathtools}
	\usepackage{amsmath}	
	\usepackage{graphicx,wrapfig,lipsum}
\hypersetup{
	colorlinks   = false,
	linkcolor=red,          
	citecolor=green,        
	filecolor=magenta,      
	urlcolor=cyan 
}
\usepackage{geometry}
\usepackage[dvipsnames]{xcolor}
 
\usepackage{xr}
\usepackage{natbib}

\begin{document}
	\title{Experimental investigations on geometry modulated solute mixing in viscoelastic media}
	\author{Bimalendu Mahapatra}
	\author{Aditya Bandopadhyay}%
	\email{aditya@mech.iitkgp.ac.in}
	\affiliation{Department of Mechanical Engineering, Indian Institute of Technology Kharagpur, Kharagpur - 721302, West Bengal, India}%
	
	\begin{abstract}
In this study, geometrically modified microchannels fabricated using stereolithography technique are employed to analyze micromixing of polymeric solutions. Experimental and numerical analyses were conducted to evaluate the qualitative and quantitative validity of the Newtonian fluid flow inside the geometrically modified channels. An in-house image processing code developed in MATLAB was used to analyze the dye concentration distributions resulting from complex fluid transport and mixing within the modified channels. Our analysis illustrates the existence of inertio-elastic instability in the geometrically modified microchannels for the transport of viscoelastic fluids. In addition, the presence of inertio-elastic instability significantly enhances mixing efficiency as a result of the interaction between viscoelasticity and modified channel geometry. 
This analysis provides important physical insights into the cost-effective design and operation of microfluidic devices that handle viscoelastic fluids and could be useful in designing and analyzing a passive micromixers that transport bio/polymeric fluids inside microchannels.
	\end{abstract}
	\maketitle
	\section{Introduction}	
	Efficient mixing is central to many applications, including process intensification, enhanced oil recovery, chemical synthesis, and micro/nanoparticle production \cite{yuan2018recent,hartman2009microchemical,sun2020three,raza2019unbalanced,mahapatra2022efficacy,hu2021design}. Mixing in microfluidics \cite{frommelt2008microfluidic,mitra2019electroosmosis,guo2019mixing,kurnia2019performance} is also integral to lab-on-a-chip devices such as blood and saliva analysis \cite{helton2007interfacial} and blood plasma mixing \cite{vandelinder2006separation}. The microconfined mixing process can be particularly challenging because the flow rates in the microchannels are usually very low and the fluids in the microchannels flow in a laminar regime with a low Reynolds number, which means that fluid flow in parallel layers without disruption between layers and mixing of the fluids relies mainly on diffusion with a very low mixing efficiency. Mixing achieved by chaotic advection generally has a higher efficiency than mixing that is purely caused by diffusion. At the macroscale, it is fairly easy to directly increase the Reynolds number in order to generate hydrodynamic turbulence. Nonetheless, at the microscale, the Reynolds number is very small due to the small characteristic length, resulting in a dominant laminar flow regime, leading to a molecular diffusion-dominated mixing mode \cite{hong2016inertio,mahapatra2022effect,johnson2002rapid}. A process that involves lateral fluid motion can enhance mixing efficiency compared to diffusion-only processes \cite{johnson2002rapid}. However, no lateral fluid motion is generated under laminar flow in a straight microchannel. 
	
	An active component, such as a micro-magnetic stirrer, can be used to create lateral fluid motion (or convective transverse flow) to enhance mixing \cite{mensing2004externally}. However, the active method requires additional processes for installing active components within a microchannel and an external field generator like a magnetic field for generating the active field \cite{mensing2004externally,mahapatra2021microconfined}. In this regard, passive micro-mixers, which do not rely on external fields, have been attracting a great deal of attention for applications that can be broadly classified into two major groups: geometric structure-guided mixing and nonlinear flow-driven mixing based on inertial or viscoelastic flow. Modulating the channel structure yields the transverse velocity component in geometric structure-guided mixing \cite{lee2011microfluidic}. Despite this, this type of mixing requires a sophisticated channel design and complicated fabrication processes. 
	
	On the other hand, flow-driven mixing especially for polymeric fluids, which is based on a viscoelastic or inertial flow of polymer solution, has obvious advantages since it produces efficient mixing without the need for complicated channels. The viscoelastic fluids have unique flow dynamics that permit flow instability to develop even when the Reynolds number is negligibly small \cite{groisman2000elastic,groisman2001efficient}. It is possible to mix viscous fluid streams by exploiting the flow instability generated in viscoelastic fluids. Many interesting phenomena can be observed as a result of the viscoelasticity of polymer solutions, such as rod climbing \cite{bird1987dynamics} and flow instability \cite{larson1992instabilities} at a very small Reynolds number. For such flows, hoop stress produces elastic turbulence along the streamlines, which has been harnessed to mix fluid streams \cite{groisman2001efficient}. According to \citet{gan2006polymer}, in viscoelastic contraction flow, the flow instabilities can cause two fluid streams with different viscosities to be mixed. A detailed analysis of the mixing of viscoelastic fluid in the microfluidic system showed a higher mixing degree than the mixing of Newtonian fluid \cite{zhang2019comparison}. In viscoelastic microfluidic flow, the enhanced mixing occurs because of the elastic stress in the fluid, which does not occur in Newtonian flow even at high flow rates \cite{hong2016inertio}.	
	
	In an attempt to examine the nonlinear flow-driven mixing based on viscoelastic flow through geometric guided structures, it is essential first to fabricate geometrically modified microchannels in an efficient way. In general, the microfluidic devices \cite{mukherjee2019single,hou2013isolation,lee2013label} are manufactured by casting polydimethylsiloxane (PDMS) into a master mould, which is made through either standard manufacturing techniques (silicon etching or SU8 lithography) or by conventional micro-milling of aluminium or polymethyl methacrylate (PMMA) sheets  \cite{hou2013isolation,warkiani2014slanted,zhou2013fundamentals}. This approach has traditionally been used to develop most of these devices. However, difficulty to build non-orthogonal and non-planar structures and its cost and labour intensiveness have limited their widespread applications and commercialization \cite{razavi2019rapid,liao2010three}. Recent developments in additive manufacturing have made it possible to fabricate functional microfluidic systems out of a variety of polymeric materials \cite{naderi2019digital}. With this unique technology, investigators can create microstructures in a very short time frame with complex shapes and geometries \cite{shallan2014cost,bhattacharjee2016upcoming}. Despite this, because the fabricated channel lacked adequate transparency, imaging (fluorescent or bright field) was not feasible through the channel. Furthermore, the removal of the residuals from channels is inconvenient for microfluidic applications that involve small cells or particles as the channel width is typically in the order of millimetres. A combination of softlithography and 3D printing of sacrificial moulds has gained significant attention lately due to its simplicity, and cost-effectiveness \cite{raoufi2020fabrication}. By fabricating microchannels with unconventional cross-sections, \citet{tang2019elasto} studied the effects of geometry on elasto-inertial focusing using the fused deposition modelling (FDM) printer. Even though this approach is suitable for fabricating microchannels with different cross-sections, FDM printing is limited in its capacity to yield high-resolution printed parts. Although stereolithography (SLA) and digital light processing (DLP) offer the potential of directly fabricating microchannels, these devices typically work in channels just a few micrometers in diameter, which makes it difficult to remove resin remnants.
	
	In order to address these deficiencies of the conventional fabrication procedures, we have followed a method of microchannel fabrication wherein open channels are printed with SLA 3D printing and are then bonded with optically transparent acrylic sheets using double-coated pressure-sensitive adhesive tape, providing a leakage-free interface for microfluidic applications. Due to its low cost, the ability to manufacture on a large scale, its user-friendliness for users who are unfamiliar with complex photolithography procedures, and the short fabrication time, this microfabrication method has gained great attention recently \cite{lee2022three,razavi20203d,ding2022giardia,preetam2022emergence,balakrishnan20203d,zargaryan2020hybrid}. Most of the above studies used Newtonian fluid in their analysis to mimic the biofluid transport. Currently, there are no studies on the transport of viscoelastic fluid/polymeric solutions through 3D printed resin channels to the best of our knowledge. Based on the aforementioned motivation, in this study geometric modulations of the microchannel and nonlinear flow-driven mixing based on viscoelastic fluid flow have been investigated together to determine mixing characteristics.
	
	\section{Experimental Methodology} \label{sec:ExpM}
\begin{figure}[ht]
		\centering
		\includegraphics[scale=0.32]{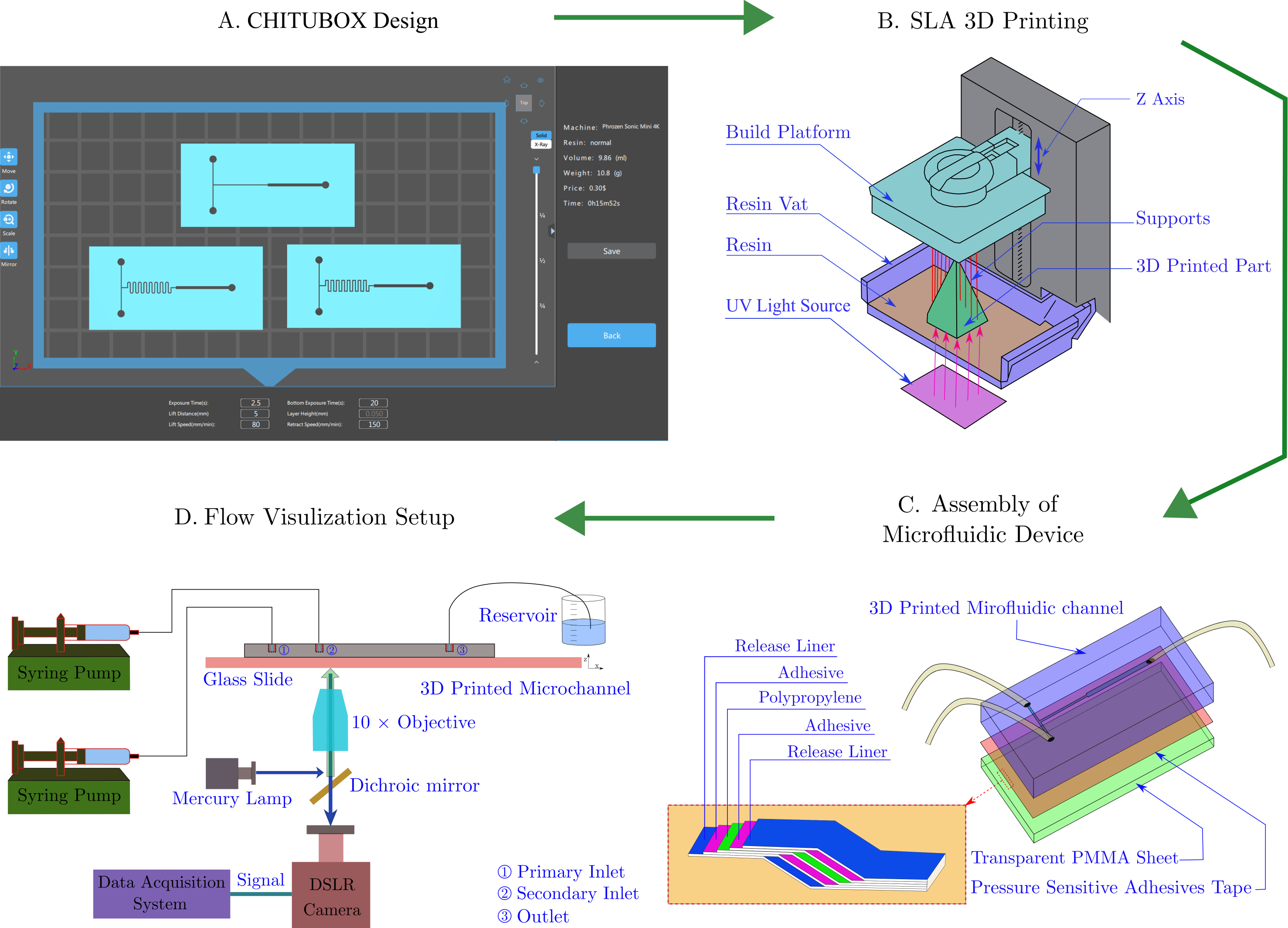}
		\caption{(Color online) The proposed workflow for fabricating microfluidic devices is pictured in the schematic. (A) Design and setting of printing parameters in Chitubox V 1.8.1. (B) 3D printing of the channel geometry was performed using a high-resolution SLA printer. (C) An adhesive tape with a double-coated pressure-sensitive coating was used to bond the part to the PMMA sheet after the 3D printed part was cleaned with isopropanol. (D) The flow visualization inside the microchannels using microscopy can be accomplished from the bottom of the channel by taking advantage of the transparency of PMMA}
		\label{Fig:1}
	\end{figure}

	In this manuscript, we have investigated the mixing characteristics for the flow of aqueous polyethylene oxide in three different microchannel geometries, namely straight, curvilinear and serpentine channels fabricated by stereolithography (SLA). All the channels have width  $w = 250$ $\mu$m and height $h = 200$ $\mu$m at the mixing region. The experiments were carried out for varying PEO concentrations using the dye tracking method and image processing analysis. The concentration distributions and mixing efficiency for all the experimental cases were quantitatively obtained from the image processing using in-house Matlab code. A schematic of the workflow from the design of the microchannels, fabrication of open channel using 3D printing, assembly of the microfluidic device, and flow visualization setup is illustrated in Fig. \ref{Fig:1}. 
	
	\subsection{Fabrication of the microchannel}
	The present study uses stereolithography for fabricating microfluidic devices \cite{song20203d,yadav2017review}. Utilizing a resin 3D printer with high resolution, the desired microchannel is printed so that the design pattern exists on the outer face and the base is attached to the build plate. This method is of particular importance for fabricating microfluidic devices as it allows the calculation of cross-sectional changes. A channel must be fabricated with the appropriate and accurate dimensions by optimally optimizing the printing parameters. When it comes to creating a high-quality channel with a great surface finish, factors like slice thickness, curing time, and total thickness of the part matter the most. A substrate with adequate optical transparency and rigidity will be needed for subsequent testing of the fluidic network made from 3D-printed microchannels. 
	The microfluidic devices were fabricated using a high-resolution SLA 3D printer (Phrozen Sonic Mini 4K) offering a resolution of 722 PPI and featuringa a 35 $\mu$m XY resolution and a 6.1$^{\prime\prime}$ monochrome LCD screen, which allows the 3D printer to print up to 4 times faster than traditional 3D printers. The desired microfluidic device is first designed using a CAD program (SolidWorks 2018) and then translated into STL format, which is suitable for printing on 3D printers. In the next step, the part design is sliced in the Z-direction using Chitubox software (Chitubox V 1.8.1).	The slicing thickness in the Z-direction or layer height is 50 $\mu$m, and the exposure time is set to 2.5 seconds. The sliced file is then sent to the 3D printer with a UV wavelength of 405 nm. Through a transparent teflon film, UV light is projected from the bottom of the resin vat. After a layer has been cured by UV light, the Z-stepper motor moves one slice upward, and the next layer begins polymerizing. This process continuous until all slices are exposed and cured. After the part is removed from the build platform, a thorough rinse and wash with isopropanol is performed, followed by an air-dry by a nozzle of compressed air. To finish the post-curing process, the microchannel is then exposed to ultraviolet light with a wavelength of 405$\pm$5 nm using a UV torch.  Inlet and outlet ports are inbuilt into the model itself and need not be punched, which is the case for softlithography. As the most promising and reproducible method, in this work, we have used double-coated adhesive tape to permanently bond 3D-printed channels to a PMMA sheet. To enhance the transparency of the optical microscopy, we have used a clear resin (ANYCUBIC clear resin) to fabricate the 3D printed parts. A transparent double-coated pressure-sensitive adhesive tape (FLEXCON) having 45 $\mu$m clear polyester film thickness was cut with a similar size to the PMMA sheet. The 3D-printed part was manually placed over one side of the pressure-sensitive-adhesive (PSA) tape, and the PMMA sheet was attached to the other side of the PSA tape. In the next step, the 3D printed part, the tape and the PMMA sheet are pressed together until no bubbles are visible at the interface between them. The Tygon tubes were eventually connected to the microchannel, and priming was performed to verify the appropriate bonding of the 3D printed part with PSA tape.
	
\begin{figure}[ht]
		\centering
		\includegraphics[scale=0.4]{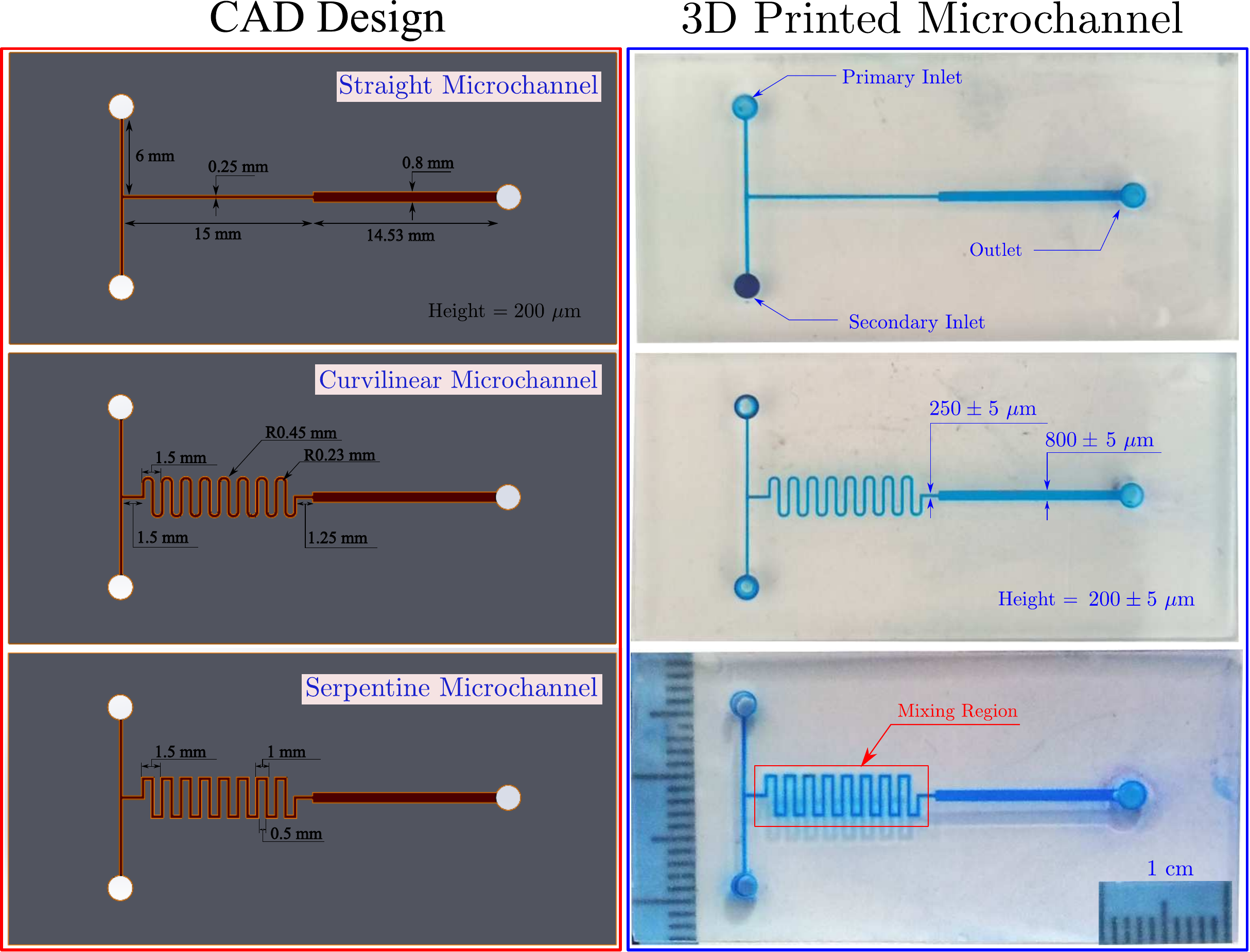}
		\caption{(Color online) A qualitative comparison between the CAD design of the microchannels and actual fabricated microchannels using SLA 3D printing is illustrated}
		\label{Fig:3}
	\end{figure}

\begin{figure}[ht]
		\centering
		\includegraphics[scale=0.4]{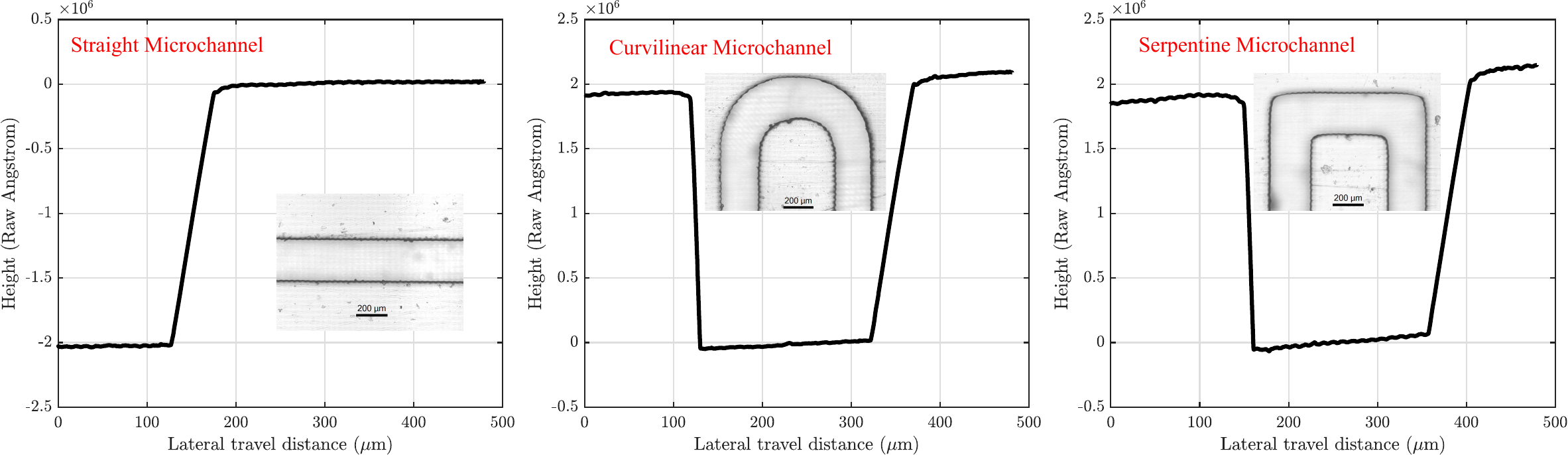}
		\caption{(Color online) Measured channel heights for straight, curvilinear and serpentine microchannels using a Veeco DEKTAK 150 profilometer and the insets illustrate channel widths measured in the OLYMPUS IX71 microscope using ProgRes CapturePro 2.9.0.1 software}
		\label{Fig:25}
	\end{figure}

	The user-friendliness of this approach appeals to a wide range of communities (e.g., biologists and chemists) without the requirement of softlithography and microfabrication facilities. A microfluidic device can be printed, fabricated, and then tested in less than 2 hours, showing the versatility of this method. Compared to the PDMS-made devices, the devices made with this technique do not suffer from deformation and leakage, making them good candidates for studying new physics, especially at relatively high Reynolds numbers. The present fabrication method is more rapid and can be developed using a low-cost raw material, especially in areas with limited resources, since a complicated microfluidic device typically costs money, time, and effort to develop. There are some inherent limitations to this process, such as a minimum XY-resolution of 100 $\mu$m and a minimum Z-resolution of 50 $\mu$m, as well as undercuts in the microchannels that are observed during curing. An example of the final device fabricated with this technique is shown in Fig. \ref{Fig:3}, where the internal channels are filled with methylene blue dye for illustration purposes. The channel dimensions in the XY plane are measured in the OLYMPUS IX71 microscope using the ProgRes CapturePro 2.9.0.1 software, and in the Z plane, a Veeco DEKTAK 150 profilometer is used to measure the height of the microchannels. The stylus radius of the profilometer is 2.5 $\mu$m. The measured channel height and width for straight, curvilinear and serpentine channels are presented in Fig. \ref{Fig:25}. For all the channel geometries the averaged height of the microchanel at the mixing region is $200\pm5$ $\mu$m and the width is $250\pm5$ $\mu$m as illustrated in Fig. \ref{Fig:3} and Fig. \ref{Fig:25}.

	\subsection{Preparation of Complex Fluid}
	The complex fluid/polymer solution is prepared by dissolving enough quantity of polymer in distilled water (DI water) to get 0.2\%, 0.3\%, and 0.4\% w/v concentrations. Here, we have considered the polyethylene oxide (PEO) (Sigma-Aldrich Chemical Co.) having a molecular weight of 4$\times 10^6$ g mol$^{-1}$ as the polymer. The solution of DI water and PEO is agitated using magnetic stirring at 300 rpm at room temperature for 36 hrs. The critical concentration ($c^*$) and entanglement concentration ($c_e$) of the PEO solutions are represented by $c^*=1/[\eta]$	and $c_e=6c^*$ \cite{arnolds2010capillary}, respectively, where the intrinsic viscosity $\eta$ is obtained from Mark-Houwink-Sakurada correlation\cite{tirtaatmadja2006drop} [$\eta$]=0.072$M_w^{0.65}$. For molecular weight $M_w=4\times 10^6$ g/mol, a critical concentration value of $c^*=0.071$\% w/v and entanglement concentration of $c_e=0.426$\% w/v is thus obtained. The relaxation time, $\lambda$, in the dilute regime is estimated using the Zimm model \cite{bird1987dynamics}
	\begin{equation}\label{eq:5366}
		\lambda_z=\frac{F[\eta]M_w \eta_s}{N_A k_B T}
	\end{equation}
	where, $\lambda_z$ is the Zimm relaxation time $M_w$ is the molecular weight, $N_A$ the Avogadro number, $k_B$ the Boltzmann’s constant, $T$ the absolute temperature, $\eta_s$ the solvent viscosity, $\eta$ the intrinsic viscosity and $F$ is determined by the Riemann Zeta function, $\zeta(3\nu^*)^{-1}=\sum_{k=1}^{\infty}i/(3\nu^*)$, $\nu^*$ is fractal polymer dimension obtained from $a=3\nu^*-1$ where, $a$ is the exponent of Mark-Houwink-Sakurada correlation. The relaxation time in semi-dilute unentangled and and semi-dilute entangled regimes are represented by $\lambda_{\text{SUE}}$ and $\lambda_{\text{SE}}$, respectively, and are calculated using the correlations $\lambda_{\text{SUE}}=\lambda_z\left(\frac{c}{c^*}\right)^{\frac{2-3\nu^*}{3\nu^*-1}}$ and $\lambda_{\text{SE}}=\lambda_z\left(\frac{c}{c^*}\right)^{\frac{3-3\nu^*}{3\nu^*-1}}$ \cite{rubinstein2003polymer,del2015rheometry}. The relaxation times and concentration ratios corresponding to the chosen concentrations in the study are listed in Table-\ref{tab:Rheoprop}. The zero-shear viscosity, $\eta_0$, of the solutions given in Table-\ref{tab:Rheoprop} are obtained from \citet{varma2020universality}
	\begin{table}[ht]
		\caption{Rheological properties of the polymeric solution }
		\centering
		\setlength{\tabcolsep}{0.55em}
		\setlength{\extrarowheight}{1.2pt}
		\begin{tabular}{llllll}
			\hline\hline
			Polymer & $M_w$ (g/mol)& $c$ (\%w/v) & $c^*$ (\%w/v) & $\eta_0$ (Pa$\cdot$s) & $\lambda$ (s) \\
			\hline
			PEO&$4\times10^{6}$& 0.2 & 0.071         & 0.009                     & 0.001833          \\
			PEO & $4\times10^{6}$ & 0.3 & 0.071       & 0.015                       & 0.00228          \\
			PEO&$4\times10^{6}$& 0.4 & 0.071        & 0.018                      & 0.00263          \\
			\hline\hline      
		\end{tabular}
		\label{tab:Rheoprop}
	\end{table}

	\subsection{Flow visualization}	
	\begin{figure}[ht]
		\centering
		\includegraphics[scale=0.75]{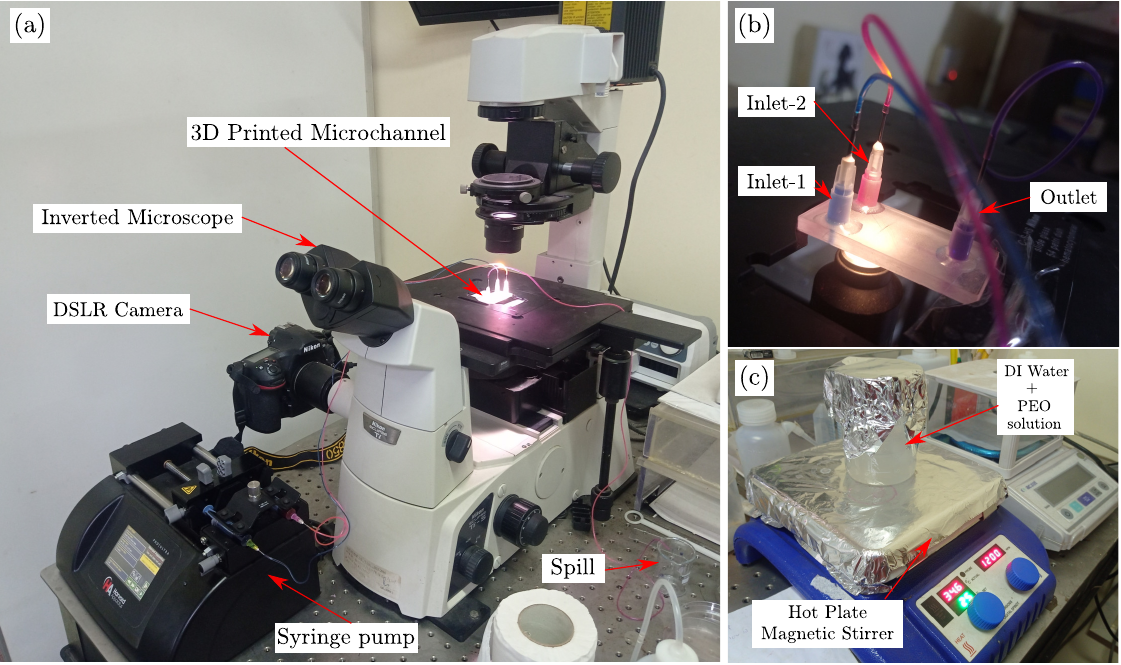}
		\caption{(Color online) (a) An illustration of the entire experimental setup for complex fluid flow and flow visualization. (b) The actual intricate microfluidic device with connected inlets and outlets. The bottom of the channel is viewed in the microscope. (c) Magnetic stirring setup for the preparation of polymer solution.}
		\label{Fig:4}
	\end{figure}
	The fluid streams from the two inlets were visualized by adding different coloured dyes to the polymeric solutions. We have added methylene blue and rhodamine-B in a 1:100 w/v ratio to the polymeric solution, which is sufficient to visualize the flow without significantly changing the viscosity of the solutions. The flow is controlled by a syringe pump (Harvard Apparatus PHD ULTRA$^{\text{\tiny{TM}}}$) quipped with two identical 5 ml syringes, and the flow rate is maintained equal for both the inlets. After the flow achieves a steady state, videos (60 fps) are captured using a digital single-lens reflex (DSLR-Nikon D850) camera installed on an inverted microscope (Nikon Eclipse T$i$). From the recorded videos, the images (1920 $\times$ 1080)  were captured and analyzed using an in-house image processing code in Matlab. Approximately five percent of the channel width near each wall was excluded because of optical interference caused by varying materials in fluid and channel walls. The complete setup used for the experiment is illustrated in Fig. \ref{Fig:4}.	
	\section{Results and Discussion}	 
	\subsection{Comparison of Experimental and Numerical results} \label{subsec:CompEN}
	Before analyzing the mixing characteristics of viscoelastic fluid flows, qualitative and quantitative validation of the experimental results is necessary. To validate, we have carried out experiments and numerical simulations on Newtonian fluid flow through the microchannels of various types, namely straight, curvilinear and serpentine channels. The Reynolds number is maintained at $Re=5$ for all three cases, and the flow rate is kept constant in both primary and secondary inlets. 
	
\begin{figure}[ht]
		\centering
		\includegraphics[scale=0.34]{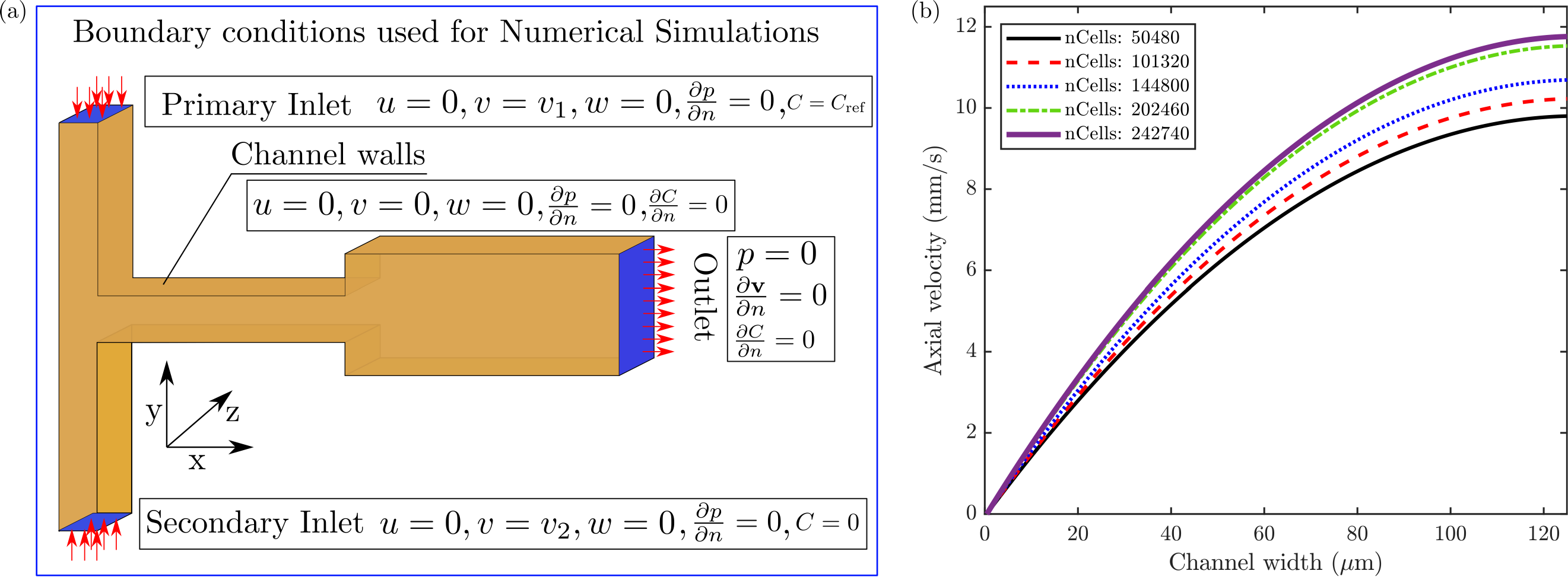}
		\caption{(Color online) (a) Schematic highlighting all the boundary conditions used in the numerical simulation. (b) Comparison of the axial velocity profile at the outlet of the mixing region for Newtonian fluid flow of serpentine channel a $Re=1$ for varying mesh resolution.}
		\label{Fig:13}
	\end{figure}

\begin{figure}[ht]
		\centering
		\includegraphics[scale=0.94]{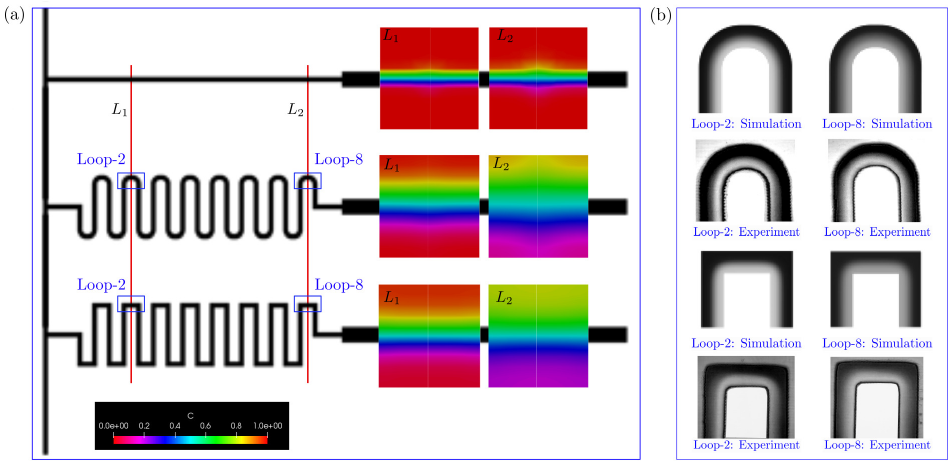}
		\caption{(Color online) Qualitative comparison of concentration profiles for experimental and Numerical and experimental data for the flow of Newtonian fluids for Re=$5$ for straight, curvilinear and serpentine channels}
		\label{Fig:5}
	\end{figure}

	\paragraph{Numerical Simulations:}
	A 3D numerical simulation of the exact flow condition is carried out using a finite volume method in OpenFoam. For a Newtonian fluid flow, where the flow is assumed to be incompressible, the governing equations of continuity and momentum at steady state are written as:
	\begin{equation}\label{eq:1}
		\nabla \cdot \mathbf{v}=0
	\end{equation}
	\begin{equation}\label{eq:2}
		\rho(\mathbf{v}\cdot\nabla\mathbf{v})=-\mathbf{\nabla} p+\nabla \cdot \mathbf{\overline{\overline{\mathbf{\tau}}}}~;~\quad \text{and} ~\quad \overline{\overline{\mathbf{\tau}}}=2\eta\overline{\overline{\textbf{D}}}
	\end{equation}
	where $\overline{\overline{\textbf{D}}}$ is the deformation rate tensor. We have used the rheoFoam solver to solve the constitutive equations Eq.\ref{eq:1},\ref{eq:2} along with the diffusive/convective transport of the neutral solute described by:
	\begin{equation}\label{eq:2562}
		\frac{\partial C}{\partial t}+\mathbf{v}\cdot\nabla C=D_c\nabla^2C
	\end{equation}
	where $C$ is the species concentration and $D_c$ is the diffusion coefficient of the neutral solute. The governing equations are integrated over Cartesian cells in the domain of computation employing their corresponding boundary conditions shown in Fig. \ref{Fig:13}(a). The interface variables are calculated by using a linear interpolation scheme. The time discretization is obtained by using the Euler scheme. The Gauss linear scheme is employed to calculate the gradient of pressure and velocity, and the divergence and Laplacian are calculated using the Gaussian linear corrected scheme. The discretized equations are computed employing the SIMPLE (\textit{Semi-Implicit Method for Pressure Linked Equations}) algorithm \cite{patankar1980numerical}, in which the discretized equations are solved through a series of cyclic guesses and correct iterations. For the iteration processes we have set the convergence criterion as $\max\limits_{r,s}$ $\vert\Upsilon^{K+1}_{r,s}-\Upsilon^{K}_{r,s}\vert\leq 10^{-6}$, where the subscripts $r,s$ denote the cell index and $K$ is the iteration index with $\Upsilon=\mathbf{v}$, $p$, and $C$.
	\paragraph{Mesh independent study:}
	In numerical simulations involving fluid flow, mesh quality plays an important role in accurately predicting flow physics. In the present numerical framework, we have considered five different mesh resolutions by varying the total number of cells ($nCells$) in the computational domain. Individually employing all the five mesh resolutions, we have compared the parametric variations inside the microchannel for varying channel geometry and Reynolds number. The axial velocity variations at the outlet of the mixing region with changing mesh resolutions for a Newtonian fluid flow at $Re=1$ through the serpentine channel are illustrated in Fig. \ref{Fig:13}(b). The grid in-dependency study shows that the solver under-predicts the flow velocity in the microchannel with coarse mesh. As the mesh resolution increases, the error in the computed flow velocity reduces. In comparison to the highest mesh resolution i.e. $nCells$ = 242740, by considering fine mesh i.e. $nCells$ = 202460, we have observed an approximate 2\% deviation in the maximum axial velocity. Consequently, we have used the fine mesh $nCells$ = 202460 for all the numerical simulations without sacrificing the accuracy of the obtained results.
	\paragraph{Qualitative comparison of dye concentration distribution:}
	In an attempt to illustrate the dye concentration distributions for both experimental and numerical analysis of Newtonian fluid flow in straight, curvilinear and serpentine microchannels, we have presented Fig. \ref{Fig:5}. The simulation results obtained exclusively from the 3D numerical simulations are presented in Fig. \ref{Fig:5}(a), which highlights the concentration distribution along the height of the channel at two axial positions, namely $L_1$ and $L_2$. We observe that for curvilinear and serpentine channels, there exist diffused concentration profiles at the outlet of the mixing region, indicating better mixing as compared to the straight channel. The concentration distribution contours at loop-2 and loop-8 from both experimental and numerical analysis are illustrated in Fig. \ref{Fig:5}(b) indicates a better mixing at loop-8 for both curvilinear and serpentine channels. From Fig. \ref{Fig:5}(b), we can also observe a good qualitative match of the concentration distribution obtained from both experimental and numerical analysis. These qualitative results will further be analyzed using image processing to obtain quantitative comparisons between the experiment and numerical simulation.
	\begin{figure}[ht]
		\centering
		\includegraphics[scale=0.55]{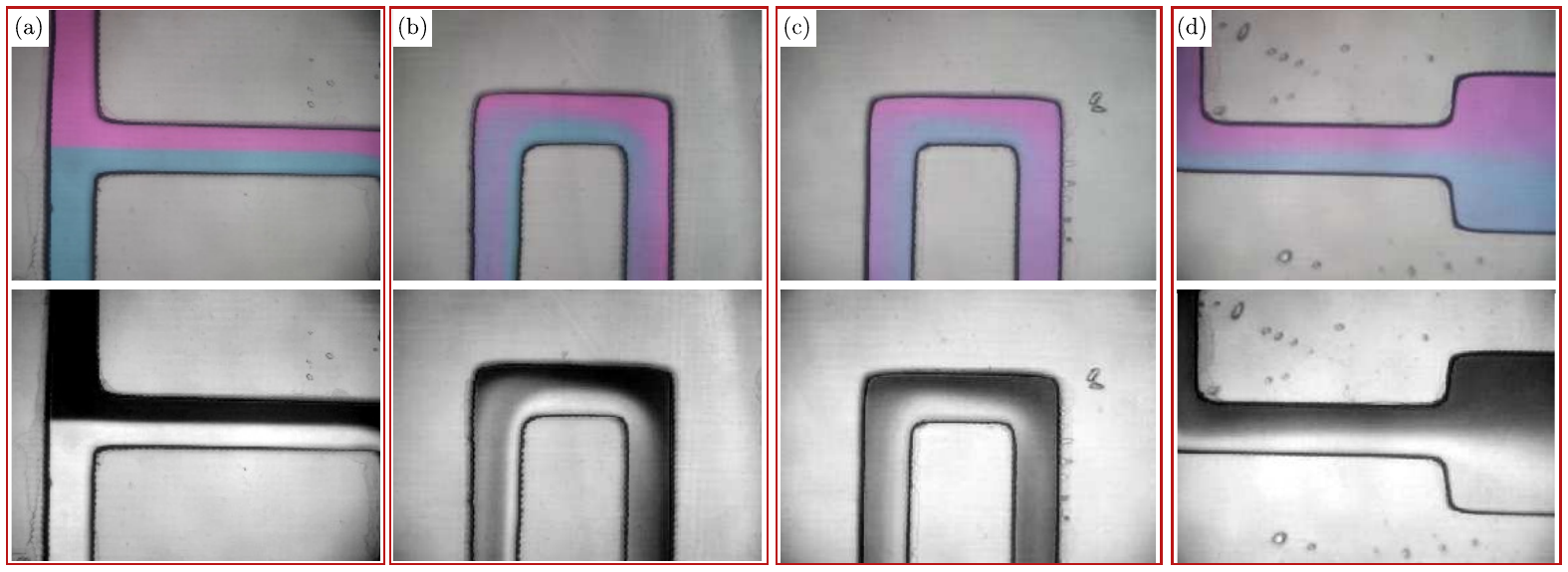}
		\caption{(Color online) Original and re-scaled images for serpentine channel $Re=10$ at (a) inlet, (b) loop-2, (c) loop-8 and at the (d) outlet of mixing region}
		\label{Fig:6}
	\end{figure}
	\paragraph{Image processing of the experimental data:}
	In order to quantify the concentration distribution, colour images at selected cross sections are converted to grey-scale images using an in-house Matlab code. In the code, first, the colour image is loaded and cropped to highlight the mixing region. The image is then converted to individual red, green and blue channels. The grey-scale image is then produced by doing arithmetic operations on the individual red, green and blue channels. For proper scaling of the obtained grey-scale image, we first calibrate the image so that at the inlet, we get a step-like distribution of concentration varying between 0 and 1. The re-scaling values of the first image (inlet) are kept constant throughout the image processing for an experimental run. In Fig. \ref{Fig:6}, we have shown the original and re-scaled images for four different cross sections. One can observe that there exists a sharp interface between the two fluid streams at the inlet (Fig. \ref{Fig:6}a) for both the original and re-scaled images. As the axial distance increases, the diffused interface is observed as a result of the fluid mixing, which is accurately captured in the grey-scale images highlighted in Fig.\ref{Fig:6}(b-d). These grey-scale images are considered for further quantitative analysis of the concentration distribution in the microchannels.
	\paragraph{Quantitative comparison of experimental and numerical results:} 
	\begin{figure}[ht]
		\centering
		\includegraphics[scale=0.38]{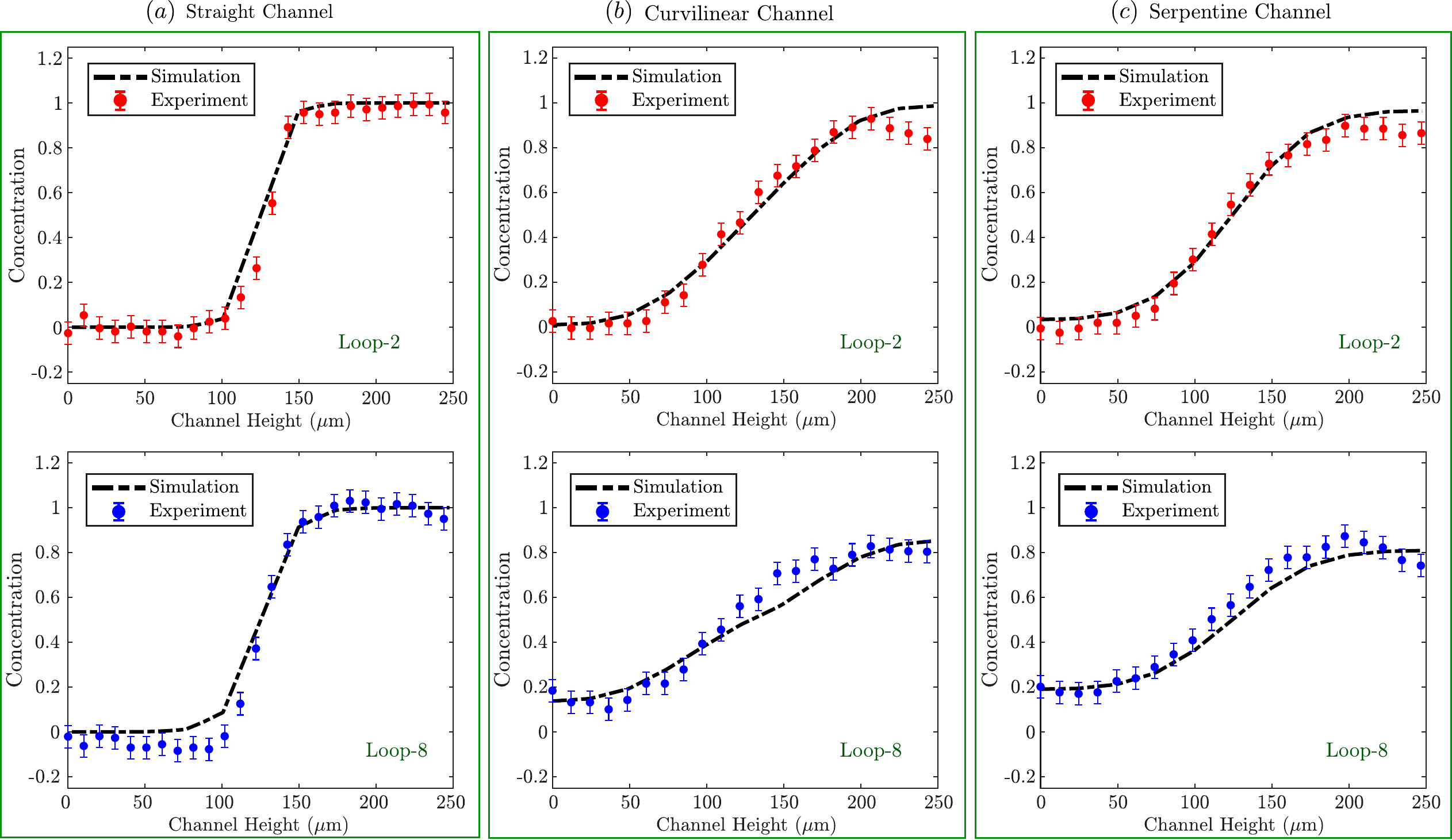}
		\caption{(Color online) Comparison of concentration profiles for numerical and experimental data for the flow of Newtonian fluids for Re=$5$ for (a) straight, (b) curvilinear channel and (c) serpentine channel.}
		\label{Fig:7}
	\end{figure}
	The grey-scale images are processed to obtain the colour intensity, which denotes the concentration distribution in the microchannel. To validate the experimental data with the numerical simulation, we have considered two axial positions, i.e. the centerline of loop-2 and loop-8 for straight, curvilinear and serpentine channels. The comparison plots for the concentration distribution considering various microchannels are presented in Fig. \ref{Fig:7}. For the straight channel at loop-2, we observe an approximate step-like distribution of the concentration distribution from both experimental and numerical analysis. The concentration distribution has an insignificant variation as we move axially from loop-2 to loop-8 in the straight channel as shown in Fig. \ref{Fig:7}(a). This indicates that the fluid streams do not show substantial mixing while in the downstream of the straight microchannel. We observe, both experimentally and numerically (refer to Fig. \ref{Fig:7}(b,c)), that the concentrations at the channel wall approach the centerline concentration for curvilinear and serpentine microchannels, indicating improved mixing characteristics when compared to the straight microchannel at the same axial position. The numerical and experimental results also indicate that at loop-8, the mixing is enhanced compared to loop-2 for curvilinear and serpentine microchannels. From the above analysis, we can conclude that the numerical and experimental results have a good agreement which validates our experimental and image processing procedures to access the concentration distribution for the flow through 3D printed microchannels.
	\subsection{Mixing efficiency calculations for viscoelastic fluid flow through 3D printed microfluidic device} \label{subsec:mixingVisco}
	Previously, we have discussed the effects of geometrical modifications to the microchannel on the concentration distribution for Newtonian fluid flows. However, the flow of viscoelastic fluids through the geometrically modified microchannels may show interesting mixing characteristics due to the interaction of modified channel geometry and complex fluid rheology, which we will investigate and analyze in this section. The viscoelasticity of polymer solutions is characterized by the Weissenberg number ($Wi=\lambda\dot\gamma_c$) where $\lambda$ is the relaxation time of the polymer solution and $\dot\gamma_c=\frac{2Q}{h^2w}$ is the characteristic strain rate \cite{rodd2007role}. The Reynolds number is defined as the ratio of inertial to viscous forces, $Re=D_h V/\nu$, where $D_h=\frac{2hw}{h+w}$ is the characteristic geometrical length, $V=\frac{Q}{hw}$ is the characteristic velocity, and $\nu=\mu/\rho$ is the kinematic viscosity. The elasticity number $El=\frac{Wi}{Re}=\frac{\eta\lambda(w+h)}{\rho h^2 w}$ is introduced to define the relative ratio of the elastic to the inertial properties of the viscoelastic flow. The P\'eclet number is defined as the ratio of the diffusion time scale to the convection time scale, i.e. $Pe=\frac{D_hV}{D}$, where $D$ is the diffusion coefficient of the dye used for visualization. The average diffusion coefficient of an aqueous solution of methylene blue/rhodamine-B is of the order $D \sim O(10^{-6})$ cm$^2$s$^{-1}$ \cite{selifonov2019determination,rani2005rapid}. For the present experimental conditions, the P\'eclet number is of the order $Pe \sim O(10^{4})$; hence the diffusive effect is not significant and can be neglected.
	\subsubsection{Viscoelastic fluid flow visualization }
	\begin{figure}[ht]
		\hspace{2em}
		\includegraphics[scale=0.65]{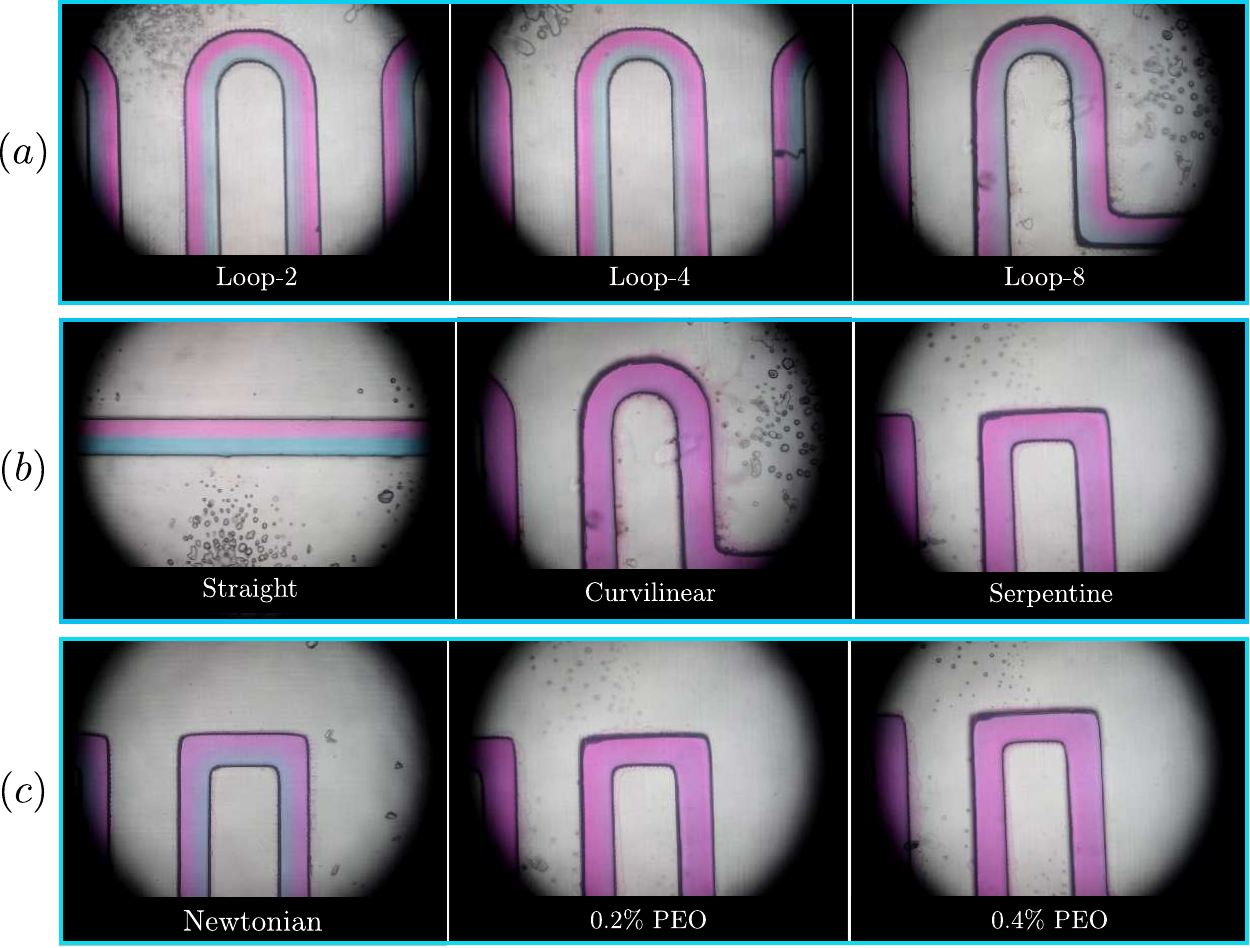}
		\caption{(Color online) Flow visualization for varying flow governing parameters (a) varying axial position in a curvilinear channel for 0.2\% PEO [$El=0.7424$], (b) for PEO concentrations of 0.3\% [$El=1.359$], straight, curvilinear and serpentine channels at Loop-8 were considered (c) three different types of fluids i.e. Newtonian [$El=0$], 0.2\% PEO [$El=0.7424$] and 0.4\% PEO [$El=2.1303$] are considered at loop-8 in a serpentine channel}
		\label{Fig:8}
	\end{figure}
	To demonstrate a qualitative comparison, in Fig. \ref{Fig:8} we present the dye distribution of the viscoelastic fluid streams coming from the two inlets for varying axial positions, modified channel geometries and changing PEO concentrations. From Fig. \ref{Fig:8}(a), we can observe that the sharp interface of the fluid streams at loop-2 becomes diffused as we move axially towards loop-8. This indicates a higher mixing as we move downstream in the microchannel. For changing channel geometry in Fig. \ref{Fig:8}(b), we observe completely diffused fluid streams for curvilinear and serpentine microchannels, whereas, for the straight channel, the fluid interface remains sharp. This shows that for the same axial positions, the mixing is augmented in modified channels as compared to the straight channel. In Fig. \ref{Fig:8}(c), we have also highlighted the comparison of dye distribution for Newtonian fluid and viscoelastic fluid by varying PEO concentrations for flow in a serpentine channel. It is observed that the two fluid streams are still visible at loop-8 for Newtonian fluid; however, for viscoelastic fluids, i.e. 0.2\% PEO and 0.4\% PEO, there is no clear distinction between the fluid streams. This indicates that an augmentation in mixing is achieved inside the microchannel using the viscoelastic fluids in comparison to the Newtonian fluid. Additionally, we have analyzed and quantified these qualitative observations and presented them in the following sections.
	\subsubsection{Procedure to evaluate mixing performance}
	In the present analysis, we have quantified the mixing performance using the probability density function of the obtained grey-scale images normalized by the average intensity value, which is similar to the procedure followed by \citet{gan2006polymer} and \citet{yang2021efficient}. Statistical analysis is conducted on the concentration distribution data obtained from image processing, which is splited in the range from 0 to 1 and divided equally into 50 bins by the discrete distribution. 
	\begin{wrapfigure}{r}{7cm}
		\vspace{-0.0cm}
		\centering
		\includegraphics[scale=0.55]{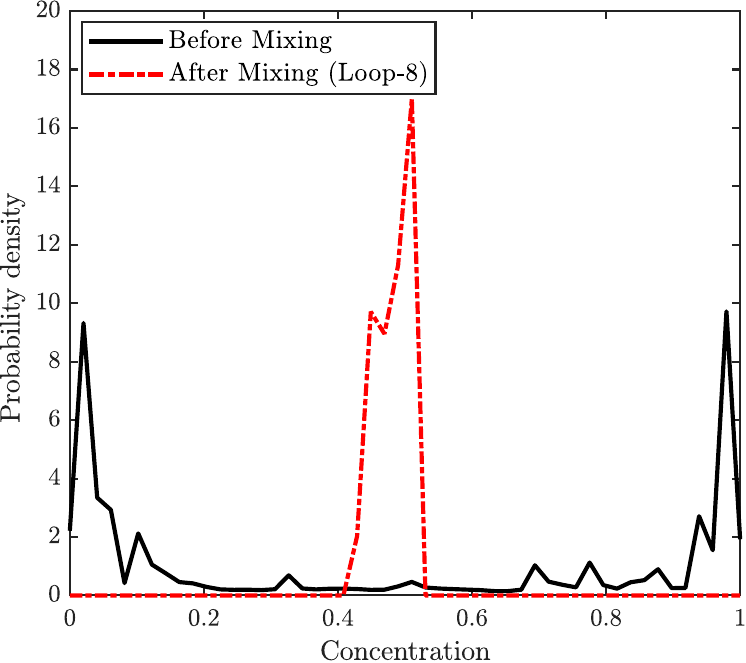}
		\caption{Probability density function (PDF) of grey-scale concentration before mixing region and after mixing region in the serpentine channel for $0.4\%$ PEO}\label{Fig:11}
	\end{wrapfigure} 
	In Fig. \ref{Fig:11}, we have highlighted the probability density function of the concentration distributions for the flow of polymeric solution ($0.4\%$ PEO) in a serpentine microchannel. There exist two significant peaks at low and high concentrations. The fact that a function shows two peaks at low and high normalized concentrations indicates that the fluids from the two inlets are not mixed adequately, while the smooth PDF gathered in the middle with nearly one peak indicates that the two streams are fully mixed. At loop-8, we observe a significant peak at the mid-range of PDF, which indicates effective mixing in that region. In the image processing procedure, the range of the grey-scale intensity is from 0 to 1. On the basis of the following function, we can calculate $\gamma_{\text{eff}}$, the mixing efficiency\cite{gan2006polymer,yang2021efficient}
	\begin{equation}
		\gamma_{\text{eff}}=\left[1-\frac{\sum_{C_i=0}^{C_i=1}\vert C_i-C_\infty \vert P(C_i)}{C_\infty}\right]\times 100\%
	\end{equation}
	where $C_i$ is the concentration data obtained from image processing, $C_\infty$ is the concentration for perfect mixing, and $P(C_i)$ is the probability density function. In the present analysis, we have maintained equal flow rates in the primary and secondary inlet; hence $C_\infty=0.5$ and for no mixing $\gamma_{\text{eff}}$ becomes 0\%, and for fully mixed fluid $\gamma_{\text{eff}}$ becomes 100\%.
	\subsubsection{Mixing characteristics of viscoelastic fluid through geometrically modified microchannels}
\begin{figure}[ht]
		\centering
		\includegraphics[scale=0.5]{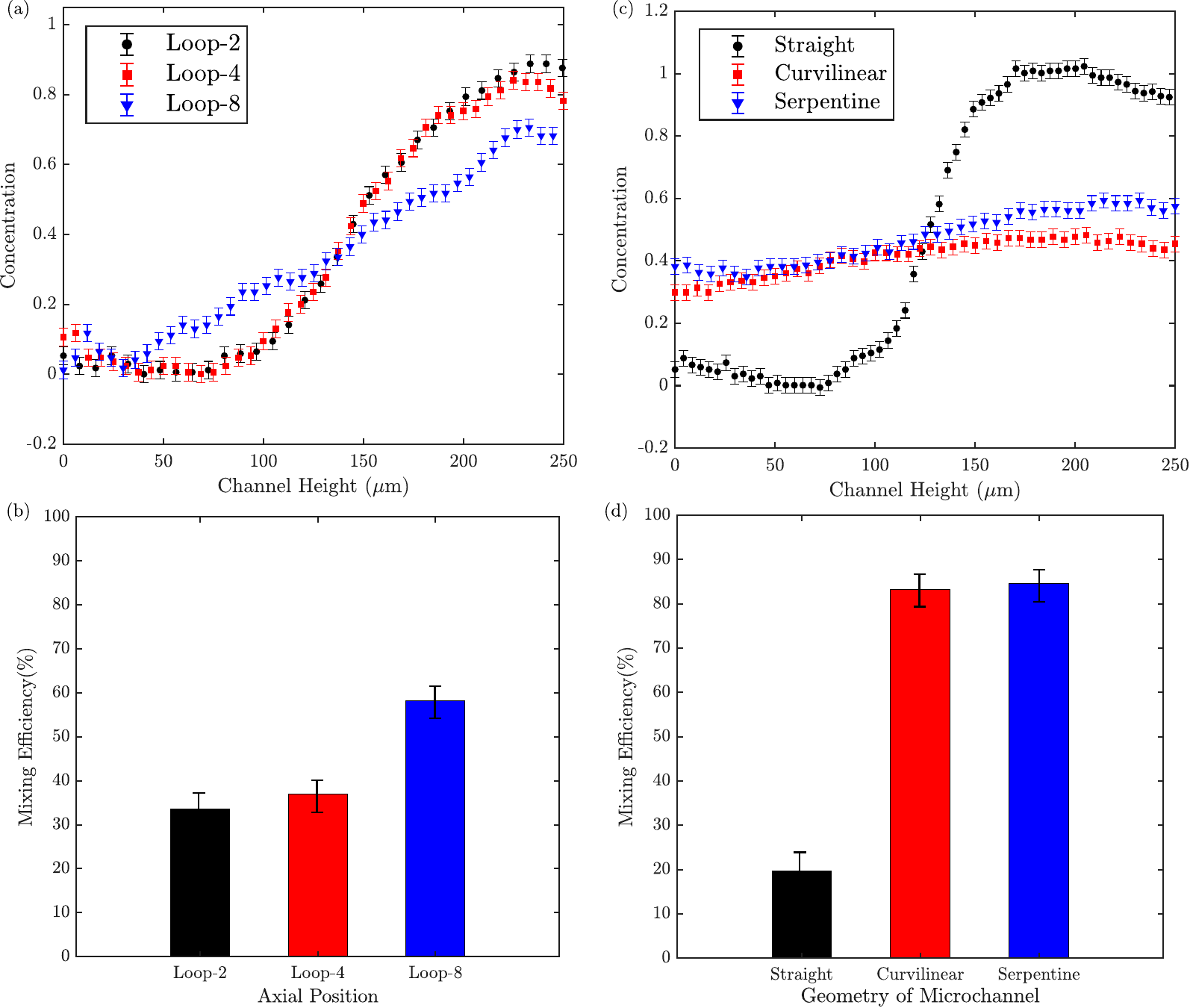}
		\caption{(Color online) (a) dye concentration distribution (b) mixing efficiency for varying axial position for a curvilinear microchannel considering 0.2\% PEO [$El=0.7424$] (c) dye concentration distribution (d) mixing efficiency for PEO concentrations of 0.3\% [$El=1.3590$] considering three channel geometries i.e. straight, curvilinear and serpentine channels at loop-8}
		\label{Fig:9}
	\end{figure}

	In previous sections, we have discussed the qualitative effects of altering channel geometry and polymer concentration on the dye concentration distribution inside microchannels. Specifically, we focus on quantifying mixing performance with varying fluid rheology and channel geometry. In Fig. \ref{Fig:9}, we highlight the variation of dye concentration and mixing efficiency with changing axial position, channel geometry and polymer concentration for the flow of aqueous polyethylene oxide inside the microchannels. We can clearly observe from Fig. \ref{Fig:9}(a) that with increasing axial distance, i.e. when we move from loop-2 to loop-8 in a curvilinear microchannel keeping $El=0.7424$, the range of concentration distribution along the channel height reduces. This is indicative of increased mixing in the downstream of the microchannel. For quantification, we have also presented a bar chart in Fig. \ref{Fig:9}(b), which highlights the enhanced mixing efficiency at loop-8 of the curvilinear microchannel. Comparing the mixing efficiency data, we obtain that there is a 10.22\% increment in mixing efficiency when we move from loop-2 to loop-4 and a 47.4\% increment in mixing efficiency when we move from loop-4 to loop-8. Increasing axial distance increases mixing efficiency, despite the absence of diffusion ($Pe\gg1$), because fluid streams encounter more channel geometry undulation moving downstream in the microchannels, which increases interaction between them.  
	
	The next step is to investigate the mixing characteristics of the viscoelastic fluid flowing through microchannels of varying geometry. During our experiments, we considered straight, curvilinear, and serpentine microchannels through which aqueous polyethylene oxide was transported at an elasticity number of $1.359$. The dye concentration distributions for varying channel geometry presented in Fig. \ref{Fig:9}(c) indicate that for curvilinear and serpentine channels, the concentration at both the walls approaches the centre-line concentration, which suggests better mixing. The straight channel, however, shows a wide deviation between near-wall and centre-line concentrations, dictating a lesser mixing than the modified channels. This is also evident from the mixing efficiency variations presented in Fig. \ref{Fig:9}(d) that a great augmentation in $\gamma_{\text{eff}}$ is achieved by modifying the channel geometry. For the curvilinear and serpentine channels, we have obtained $\gamma_{\text{eff}}$ = 83.2\% and 84.6\%, respectively, which is significantly higher than the mixing efficiency obtained for the straight channel ($\gamma_{\text{eff}}$ = 19.6\%). This augmentation in $\gamma_{\text{eff}}$ for the modified channel as compared to the straight channels is because of the presence of the large number of channel bents which facilitate the viscoelastic fluid mixing due to the generation of transverse velocities at the bents because of the interaction between two fluid streams. Previously, \citet{kim2017inertio} have shown the occurrence of flow instability even at a low Reynolds number in viscoelastic fluid flow for a microchannel having a 90$^\circ$ bent geometry. In the present study, the curvilinear/serpentine channels exhibit several bents that may contribute to generation of inertio-elastic instability which in turn enhances the mixing efficiency.\\
	\begin{figure}[ht]
		\centering
		\includegraphics[scale=0.35]{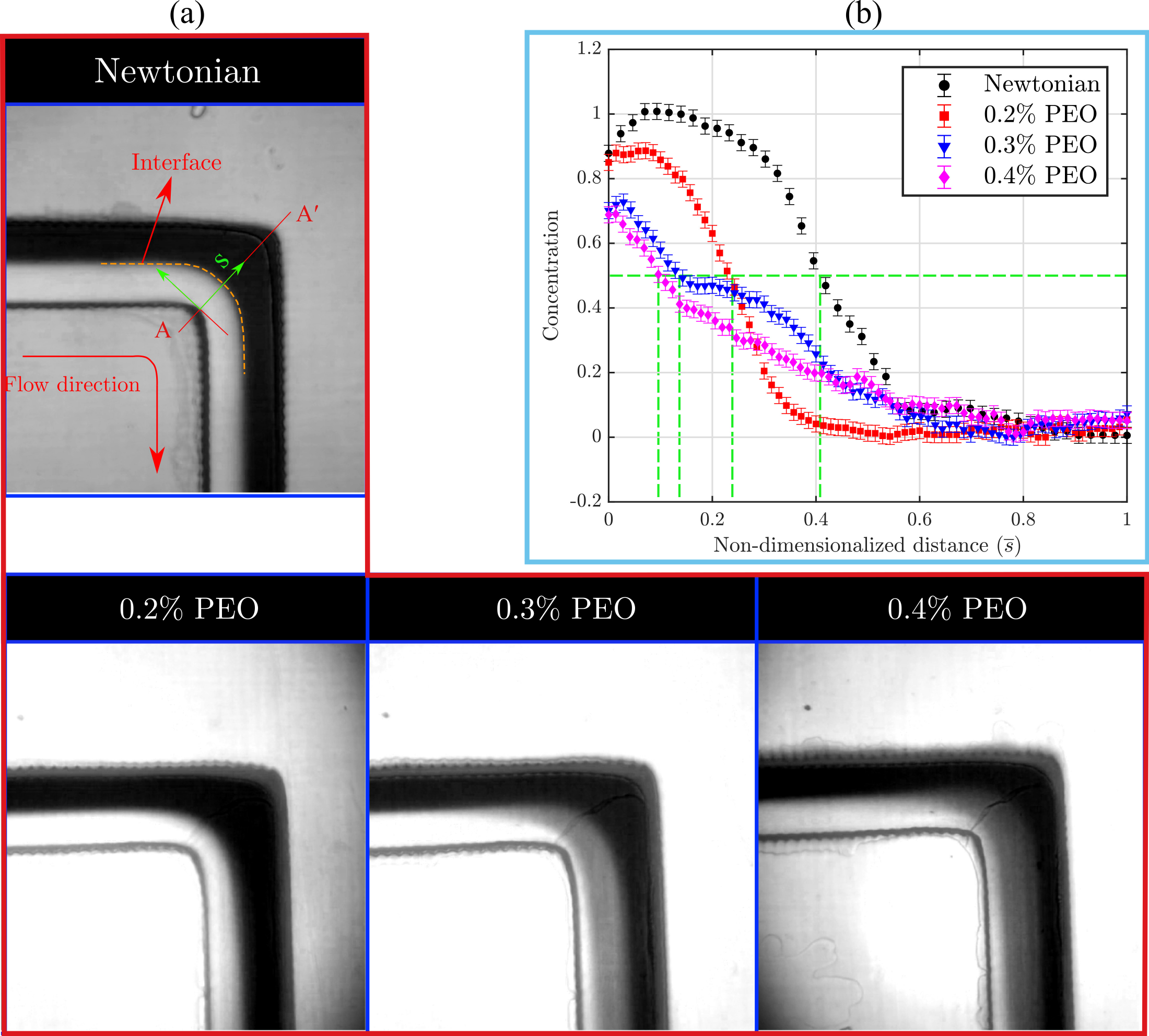}
		\caption{(Color online) (a) Figure highlights the augmentation of diffused interface or flow instability of the viscoelastic fluids with increasing PEO concentration at the 1$^{st}$ $90^\circ$ bent, (b) dye concentration distribution along a non-dimensionalized distance $\overline s$ at the channel bent i.e., $\text{AA}^\prime$ for varying PEO concentration. }
		\label{Fig:10}
	\end{figure}
	
	It is essential at this point to analyze the effect of polymeric concentration on the flow behaviour and mixing characteristics inside the 3D printed microchannels. Alteration in the polymer concentration significantly changes the rheological properties of the polymeric solution, which is expected to affect the fluid mixing considerably. Hence, in our experiments, we have considered the transport of DI water (Newtonian) and three different polymeric solutions (viscoelastic) in a serpentine microchannel. To investigate whether there exits inertio-elastic instability in the present experimental conditions, we focus on the steady flow of the two fluid streams at the 1$^{st}$ $90^\circ$ bent before the mixing region in the microchannel. We observe that (refer to Fig. \ref{Fig:10}(a)), there exists a sharp interface between the two fluid streams for the flow of DI water. Nevertheless, the fluid interface becomes diffuse as polymeric concentrations increase, and the diffused interface widens as well. 
	
	In order to quantify the observation, we have considered a cut-section $\text{AA}^\prime$ at the channel bent and obtained the dye concentration distribution along a non-dimensionalized distance `$\overline s$' for varying polymer concentration. We attempt to assess the widening of the interface by tracking the point of dye concentration $\overline c \approx 0.5$ along $\text{AA}^\prime$, we denote this concentration as $\overline c_m$ i.e., mean concentration. From Fig. \ref{Fig:10}(b), it is clear that for Newtonian fluid (DI water) flow, mean concentration $\overline c_m$ occurs at $\overline s \approx 0.41$. As the polymeric concentration is increased to 0.2\% PEO, mean concentration $\overline c_m$ shifts towards a lower $\overline s$ value and $\overline c \approx 0.5$ occurs at $\overline s \approx 0.23$. With further increase in PEO concentration, we obtain $\overline s \approx 0.095$ and $\overline s \approx 0.115$ at which the mean concentration occurs for 0.3\% PEO and 0.4\% PEO, respectively. It is also essential to note that, as the PEO concentration increases the concentration at the inner corner of channel bent reduces and also there is a reduction in slope of the dye concentration distribution. The lowering of non-dimensionalized distance $\overline s$ and reduction in slope of $\overline c$ distribution along $\text{AA}^\prime$ is a conspicuous indication of widening of the diffused interface and augmentation of inertio-elastic instability.
	
	\begin{figure}[ht]
		\centering
		\includegraphics[scale=0.52]{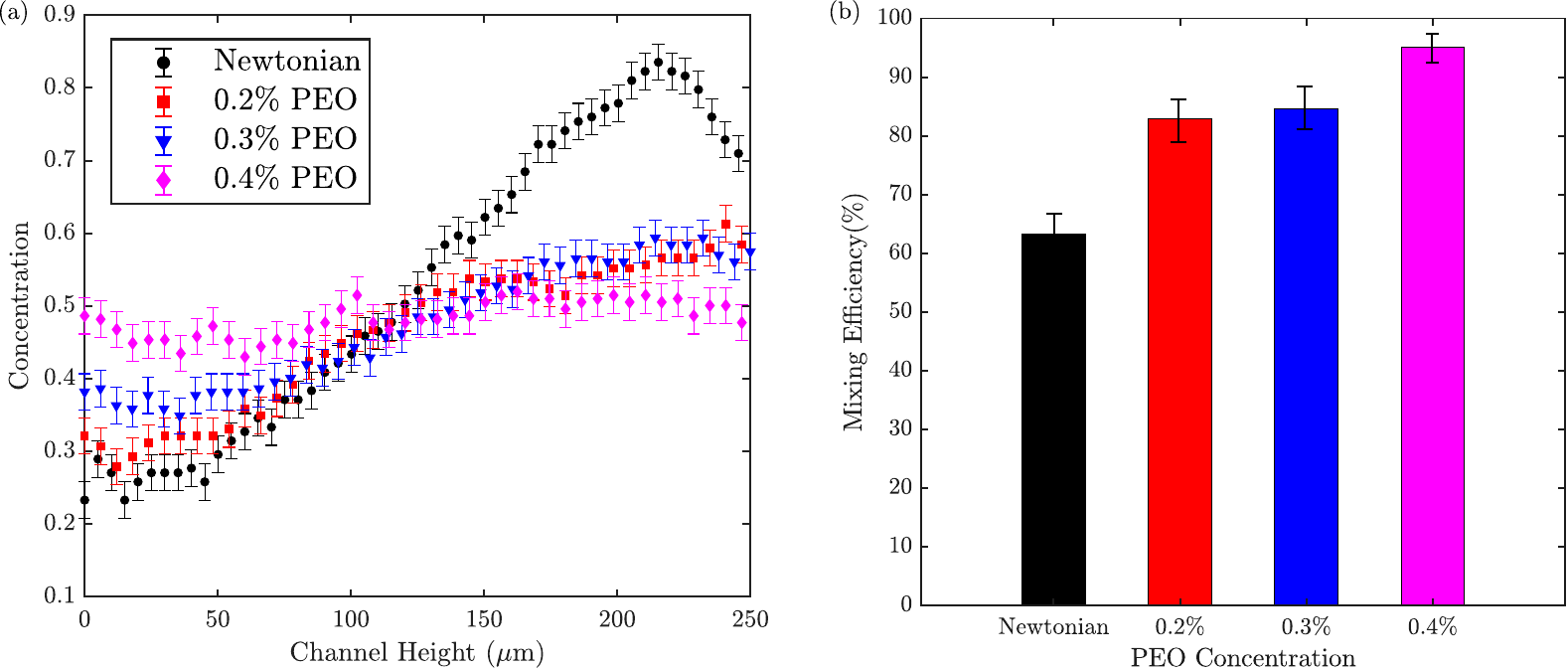}
		\caption{(Color online) (a) dye concentration distribution (b) mixing efficiency for Newtonian fluid [$El=0$] and three different PEO concentrations i.e. 0.2\% [$El=0.7424$], 0.3\% [$El=1.3590$] and 0.4\% [$El=2.1303$] are considered in a serpentine channel at loop-8}
		\label{Fig:14}
	\end{figure}
	
	The mixing characteristic will likely be altered when PEO concentration is increased due to inertio-elastic instability generated in the microchannels. Figure \ref{Fig:14}(a) highlights dye concentration distribution at loop-8 for various polymeric solutions. We observe that, for Newtonian fluid, the near-wall concentrations are far apart from the centre-line concentration; however, as we increase the polymer concentration, the dye concentrations at both the channel walls approach the centre-line concentration depicting an enhanced mixing scenario. Quantitative variation in mixing efficiency for Newtonian and viscoelastic fluids is highlighted in Fig. \ref{Fig:14}(b), which also indicates the mixing efficiency augmentation with increasing polymeric concentration. Our results demonstrate that the flow of 0.2\% PEO solution ($El=0.7424$) yielded a mixing efficiency of 83\% indicating a 31.12\% augmentation in $\gamma_{\text{eff}}$ as compared to the mixing efficiency of 63.3\% obtained for the flow of DI water ($El=0$) in the serpentine channel. An increase in the elasticity number from $El=0.7424$ to $El=1.359$ results in an increment of mixing efficiency, i.e., from 82.96\% to 84.56\%, respectively. As the elasticity number increases from $El=1.359$ to $El=2.1303$, we observed a 12.45\% improvement in mixing efficiency. The upsurge in mixing efficiency with increasing elasticity number is a reflection of increased inertio-elastic instability in the system, which intensifies with the increment in geometric undulations in the modified microchannels. 
	\section{Conclusions}\label{sec:conc}
	In the present study, we have examined the flow-driven mixing based on viscoelastic flow through geometrically modified microchannels. The microchannels are fabricated employing stereolithography technique. In order to evaluate the qualitative and quantitative validity of Newtonian fluid flow inside the geometrically modified 3D printed channels, both experimental and numerical analyses were conducted. Three dimensional numerical simulations were performed using the finite-volume based solver in OpenFoam to resolve the Newtonian fluid flow through straight, curvilinear and serpentine microchannels. Our experimental results were substantiated by the numerical simulation results, establishing the validity of our experimental procedure. Our observations of the effect of different parameters on mixing efficiency have led us to the following conclusions:
	\begin{itemize}
		\item  A geometric modification of the microchannel, combined with an increase in polymer concentration in the complex fluid, results in enhanced fluid mixing.
		\item  Our analysis illustrates the existence of inertio-elastic instability in the geometrically modified microchannels for the transport of polymeric solutions which is responsible for generating diffused fluid interface.
		\item The complex interaction between viscoelastic fluids and modified channel geometry makes inertio-elastic instability a major factor in enhancing mixing efficiency.
		\item Based on the results of quantitative analysis, the mixing efficiency in the curvilinear and serpentine channels increased by 324.48\% and 331.63\%, respectively, over the straight channels.
		\item  The viscoelastic fluid flow ($El=2.1303$) inside the serpentine channel yields a mixing efficiency of 95.26\% which is 50.48\% higher than the mixing efficiency obtained for the flow of Newtonian fluid.
	\end{itemize}
	The results of this study provide important physical insights into the cost-effective design and operation of microfluidic devices for the handling of viscoelastic fluids. These findings could be useful in the design and analysis of a passive micromixer that transports bio/polymeric fluids within microchannels efficiently.

	\section*{Conflict of interest}
	\noindent The authors declare that they have no conflict of interest.
	\section*{Data availability statement}
	\noindent The data that support the findings of this study are available from the corresponding author upon reasonable request.
	\vspace{1cm}
	\hrule
	\hrule
	\bibliographystyle{unsrtnat}
	\bibliography{3DPrinted_Microfluidics}
\end{document}